# Recent progress in atomistic simulation of electrical current DNA sequencing

*Han Seul Kim and Yong-Hoon Kim**

School of Energy, Environment, Water, and Sustaibability, Korea Advanced Institute of Science and Technology, 291 Daehak-ro, Yuseong-gu, Daejeon 305-701, Korea

* Corresponding author. Tel.: +82 42 350 1717; fax: + 82 42 350 1710.
*E-mail address*: y.h.kim@kaist.ac.kr

## Abstract

We review recent advances in the DNA sequencing based on the measurement of transverse electrical currents. Device configurations proposed in the literature are classified according to whether the molecular fingerprints appear as the major (Mode I) or perturbing (Mode II) current signals. Scanning tunneling microscope and tunneling electrode gap configurations belong to the former category, while the nanochannels with or without an embedded nanopore belong to the latter. The molecular sensing mechanisms of Modes I and II roughly correspond to the electron tunneling and electrochemical gating, respectively. Special emphasis will be given on the computer simulation studies, which have been playing a critical role in the initiation and development of the field. We also highlight low-dimensional nanomaterials such as carbon nanotubes, graphene, and graphene nanoribbons that allow the novel Mode II approach. Finally, several issues in previous computational studies are discussed, which points to future research directions toward more reliable simulation of electrical current DNA sequencing devices.

**Contents**

## 1. Introduction

The sequencing of DNA not only has significant scientific implications in the context of deciphering the fundamental code of life but also represents an enormous opportunity to improve the well-being of humankind by ushering in a new era of personal or precision medicine (Mardis 2011; Rabbani et al. 2014). A decade after the completion of the Human Genome Project in 2003, effort is now devoted into the development of next-generation DNA sequencing technologies that can meet the '$1,000 genome' goal set by National Institute of Health. In this endeavor, in contrast to the second-generation DNA sequencing technologies that still require polymerase chain reaction amplification and fluorescent labeling as in the first-generation counterpart, the newly-emerged third-generation DNA sequencing technologies propose single molecule detection based on changes in currents.

The fundamental ingredient of the third-generation DNA sequencing technology is a nanopore, through which a double-stranded or single-stranded DNA translocates in a linear conformation when it is driven by an electric field (Branton et al. 2008; Yokota et al. 2014; Zwolak and Di Ventra 2008). Nanopore-based DNA sequencing was first proposed for protein pores such as α-hemolysin (Kasianowicz et al. 1996) and MspA (Butler et al. 2008). In the biological nanopore approach, trans-membrane longitudinal-direction ionic current blockades are monitored as schematically shown in Fig. 1a. While the biological nanopore approach has advantages such as the well-defined atomic-scale pore size and corresponding good signal-to-noise ratio, it also suffers from problems such as the instability of lipid membranes in an electric field.

As an alternative to biological nanopores, solid-state nanopores have been actively investigated for the next-generation DNA sequencing. Within this scheme, in addition to the longitudinal-direction ionic current, another fundamentally different sequencing mode, i.e. the transverse-direction electrical current is available (Fig. 1b). The solid-state nanopore DNA sequencing approach based on transverse electrical currents will allow employing advanced semiconductor device fabrication techniques that are well-established in the microelectronics industry. It has also attracted much attention in that novel low-dimensional nanomaterials such as carbon nanotubes (CNTs), graphene, and graphene nanoribbons (GNRs) can be adopted as electrodes and/or nanopores, which will potentially provide novel device geometries and improved device characteristics.

The two core technologies of the transverse electrical current DNA sequencing approach are (1) reading and distinguishing nucleobases at the single-molecule level (sensing and electronics) and (2) understanding and controlling the DNA translocation dynamics (nanofluidics). Theory and computation have been playing an important role in both area by proposing various novel reading mechanisms and providing atomistic pictures of translocation processes. Indeed, the concept itself was theoretically proposed first along the line of molecular electronics that involve metal electrodes (Lee and Thundat 2005; Zwolak and Di Ventra 2005). Since then, proof-of-principles experiments using Au electrodes have been successfully carried out (Chang et al. 2010; Huang et al. 2010; Ohshiro et al. 2012; Tsutsui et al. 2010). The employment of novel non-metal low-dimensional materials such as CNTs (Chen et al. 2012; Kim et al. 2014; Meng et al. 2006; Meunier and Krstić 2008; Sadeghi et al. 2014a), graphene (Ahmed et al. 2012; Postma 2010; Prasongkit et al. 2011, 2013), GNRs (Ahmed et al. 2014b; Avdoshenko et al. 2013; Cho et al. 2011; Girdhar et al. 2013, 2014; He et al. 2011; Jeong et al. 2013; Min et al. 2011; Nelson et al. 2010; Rajan et al. 2014; Rezapour et al. 2014; Saha et al. 2011; Shenglin et al. 2014; Zhang et al. 2014; Zhao et al. 2012), and other low-dimensional materials were also first proposed by theoreticians (Amorim and Scheicher 2014; Farimani et al. 2014; He et al. 2014; Sadeghi et al. 2014a; Sadeghi et al. 2014b; Thomas et al. 2014), and along this line significant experimental advances are currently being made (Chen et al. 2013; Garaj et al. 2010; Merchant et al. 2010; Schneider et al. 2010; Traversi et al. 2013).

The objective of this review is to critically review recent progress made in the computational study of DNA sequencing based on the detection of transverse electrical currents. Referring other review articles for those on the DNA translocation dynamics and control (Fyta et al. 2011; Luan et al. 2012; Zwolak and Di Ventra 2008), we will particularly focus on the single DNA sensing mechanisms together with novel low-dimensional materials and corresponding device geometries. In Table I, we have summarized the representative theoretical literature in terms of the electrode material, device geometry, modeling features, simulation levels, signal type and ranking, and reading mechanism. The organization of this article is as follows: In Sec. 2, we first summarize the theoretical formulation for the calculation of electrical currents in nanoscale junctions. Proposed device configurations and sensing mechanisms are categorized according to whether the molecular signals arise from the tunneling current (Mode I) or electrochemical gating effect (Mode II). In the remainder, we will divide the discussion in terms of inorganic materials used as the probe in the DNA sequencer. Basic features in the electronic and transport properties of CNTs, graphene, and GNRs will be also summarized. In Sec. 3, we first consider the DNA sequencing approaches employing metal nanoelectrodes that can be used as a scanning tunneling microscope (STM) tip or nanogap electrodes (Mode I), for which proof-of-principles experiments have been successfully carried out. A big advantage of employing recently emerged low-dimensional nanomaterials such as CNTs and graphene is that it can potentially allow a novel DNA sequencing mechanism (Mode II). In Sections 4 and 5, we will review the computational studies that employed CNTs and graphene, respectively. In addition to CNTs and graphene, a whole new family of novel two-dimensional nanomaterials such as hexa-boron nitride (hBN), silicene, and transition metal dichalcogenides (e.g. $MoS_2$) have been attracting great attention, and now several studies have appeared on their applications to the DNA sequencing (Sec. 6). We will also discuss potential issues in the previous computational studies of the electrical current DNA sequencing (Sec. 7), which naturally lead us to the future research directions (Sec. 8).

## 2. Background of electrical-current DNA sequencing simulations

## 2.1 Non-equilibrium Green's function theory for quantum transport

The Landauer picture (Datta 2005; Di Ventra 2008) provides a conceptual foundation to understand the steady-state currents in meso- and nano-scale junctions. It gives the expression for the current-bias voltage *I-V* characteristics as

$$I = \frac{2e}{h}\int_{\infty}^{\infty} T(E,V)[f(E-\mu_1) - f(E-\mu_2)]dE \ , \tag{1}$$

where $\mu_1 - \mu_2 = eV$. Accordingly, in the zero-bias limit, conductance is given by

$$G \approx G_0\, T(E_F), \tag{2}$$

where $G_0 = 2e^2/h$ is the quantum of conductance, and $E_F$ is the Fermi level. Note that the Landauer formulation is based on the assumptions that the electrons in the channel as well as two electrodes are non-interacting and inelastic scattering events are negligible.

As a scheme to compute transmission, one often resorts to the non-equilibrium Green's function theory (NEGF). It is a well-established theory to describe charge transport processes in nanoscale junctions. While it is a general formalism that describes an interacting system with various many-body effects that induce inelastic (energy-non-conserving) and incoherent (phase-breaking) processes, one often resort to the non-interacting (mean-field) approximation and neglects inelastic and incoherent transport processes that results in a simple expression for the transmission (Datta 2005; Di Ventra 2008),

$$T(E,V) = \mathrm{Tr}[\Gamma_1(E,V)G(E,V)\Gamma_2(E,V)G^{+}(E,V)], \tag{3}$$

where $G(E) = (ES - H + \Sigma_1 + \Sigma_2)^{-1}$ is the retarded Green's functions, where *H* and *S* are the hamiltonian and overlap matrices, $\Sigma_{1/2} = x_{1/2}g_{S1/2}x_{1/2}^{+}$ is the self-interaction energy, $x_{1,2}$ is the molecule-electrode 1/2 contact part of the total $ES - H$ matrix, $g_{S1,2}$ is surface Green's function, and $\Gamma_{1/2} = i(\Sigma_{1/2} - \Sigma_{1/2}^{+})$ is the broadening matrix.

The single-particle (non-interacting) hamiltonian $H$ can be obtained from the first-principles density-functional theory (DFT) within the Kohn-Sham approximation or tight-binding (TB) approximations. For the accurate band alignment between two electrodes and DNA, one should treat the semi-infinite bulk nature of electrodes in an accurate manner (Kim 2008; Kim et al. 2006). In Sec. 7, we will discuss possible issues in the computational level as well as the modeling of atomic configurations.

### 2.2 Applications of electrical currents to DNA sequencing

In terms of device configurations and reading mechanisms, we categorize the theoretically proposed transverse electrical current DNA sequencing methods into two main categories:

• Mode I: The first is to translocate single-stranded DNA (ssDNA) through a nanoscale gap formed between two electrodes with negligible currents and detect major current signals arising from nucleobases when they bridge the nanogap (Fig. 1c). The molecular sensing mechanism of Mode I is thus electron tunneling.

• Mode II: The second is to use a nanochannel through which a relatively large current is already flowing and detect perturbing signals resulting from a ssDNA passes by (Fig. 1d). The reading mechanism of Mode II roughly corresponds to the electrochemical gating.

The specific realization of a DNA sequencing scheme in each mode and its additional sub-categorization critically depends on the choice of electrode materials. While the specifics will be discussed in later sections, briefly, we will group the nanogap-based Mode I into the configuration in which a fixed ssDNA is read by a moving electrode (Mode I-A) and that in which a translocating ssDNA is read by fixed electrodes (Mode I-B). Mode I-A, which is essentially the STM setup, is almost exclusively based on the electrodes made of metals such as Au or Pt. On the other hand, Mode II has been mostly proposed in the context of employing novel low-dimensional materials. Mode I-B can be realized with both conventional metal and novel carbon (and related) nanomaterials.

In Mode I, the DNA molecules bridge two nanoelectrodes separated by a finite gap, and the charge transport process will be dominated by the non-resonant tunneling. In this transport regime, assuming that the conductance at $E_F$ is dominated by a single level $E_0$ closest in energy to $E_F$ (scalar version of *H*) and the coupling strengths between $E_0$ and electrode states $\gamma_{1/2}$ (scalar version of $\Gamma_{1/2}$) are small, Eqs. (2) and (3) are reduced to:

$$G \approx G_0 \times \frac{\gamma_1 \gamma_2}{(E_F - E_0)^2}. \tag{4}$$

For the ssDNA molecules, the highest occupied molecular orbital (HOMO) levels $E_{HOMO}$ of the four nucleotides, purine deoxyadenosine 5'-monophosphate (dAMP), deoxyguanosine 5'-monophosphate (dGMP), and pyrimidine deoxycytidine 5'-monophosphate (dCMP) and deoxythymidine 5'-monophosphate (dTMP) (Fig. 2), become the determinant levels $E_0$. Note that $\gamma_{1/2}$ represent the strength of the coupling to the electrodes 1/2 (or $\gamma_{1/2}/\hbar$ can be interpreted as the rates at which an electron placed in the level $E_0$ will escape into the elctrodes 1/2) (Datta 2005). According to Eq. (4), if $\gamma_{1/2}$ are similar for the four nucleobases and the electrode-DNA coupling does not modify the ordering of $E_{HOMO}$, then the conductance signal ordering will be dGMP > dAMP > dCMP > dTMP. On the other hand, if $\gamma_{1/2}$ play a more important role than $E_{HOMO}$ and $\gamma_{1/2}$ are determined by the physical sizes of the nucleobases, then one can approximately expect the ordering of two-ring purine groups (dGMP and dAMP) > one-ring pyrimidine groups (dCMP and dTMP). In reality, these two factors are closely interrelated and particularly $\gamma_{1/2}$ can strongly depend on the nature electrodes, so producing a well-defined current signal ordering can be a formidable experimental challenge. Finally, note that the four nucleobases are distinguishable in terms of the number and arrangement of chemical functional groups ($NH_2$, C=O, and $CH_3$). Therefore, we can expect that utilizing the functional group-specific charge transfer and associated differences in chemical connectivity with electrodes $\gamma_{1/2}$ could also lead to a novel DNA sequencing mechanism (Kim et al. 2014).

### 2.3 Novel low-dimensional electrode materials

The DNA sensing schemes described in the previous section, particularly the Mode II approach is closely related with novel low-dimensional materials. We thus briefly discuss their properties in view of the electrical-current DNA sequencing.

Note that the average spacing between neighboring nucleotides in a stretched ssDNA is about 7.4 Å, and this roughly sets the maximum dimension of nanoelectrodes that can achieve the single-molecule resolution in the DNA sequencing. Together with the exponential growth of nanoscience and nanotechnology on low-dimensional carbon nanomaterials, it would be natural to consider CNTs and graphene with excellent charge transport properties (Charlier et al. 2007; Cresti et al. 2008; Das Sarma et al. 2011; Kang et al. 2009) for the DNA sequencing based on electrical currents. Moreover, both CNTs and graphene provide enormous possibilities in tailoring their electronic and transport properties by controlling their atomic structures, which represents a huge potential for biosensing applications in general (Artiles et al. 2011; Balasubramanian et al. 2012).

Being one-atom-thick semimetal with the linear dispersion in their density of states (DOS) and transmission (Fig. 3a), graphene could be used as nanoelectrodes with extremely high spatial resolution in the Mode I-B configuration. Next, by rolling up graphene into a cylindrical form, one can achieve both semiconducting and metallic CNTs. Both metallic and semiconducting CNTs can be prepared according to their chirality, and in Fig. 3b we show the DOS and transmission of metallic (5,5) CNT with the diameter of 6.78 Å.

Another one-dimensional analog of the graphene family is its nanoribbon form. By forming graphene nanoribbons (GNRs), one can introduce a finite bandgap that is necessary to achieve a transistor function (Fig. 3c). However, unlike in CNTs, we have GNR edges that provide another special degree of freedom in transport. Specifically, with the zigzag graphene nanoribbon (zGNR), one can access anomalous spin-polarized edge states that mainly determine its transport properties. This should be contrasted with the case of armchair GNR (aGNR), in which the central region becomes the main transport channel. In the ground state of pristine zGNRs, the spins along each zigzag edge are aligned in a ferromagnetic (FM) fashion whereas these two FM edge states have an antiferromagnetic (AF) ordering, which results in the zero total spin polarization. This ground state of zGNR presents the degenerate $\alpha$- and $\beta$-spin bands and as in aGNR present a finite bandgap inversely proportional to the ribbon width. However, the energetic stability of the zGNR with the fully FM spin configuration is almost comparable to that of the above-described zGNR in the ground-state AF spin configuration. Because the latter FM zGNR is metallic, it is more suitable to be used as the nanoelectrode. In Fig. 3d, we show the DOS and transmission of zGNR with the width of 1.84 nm.

Additional big advantage common to CNTs, graphene, and GNRs is the very large degree of freedom in functionalization, which can tailor their atomistic, electronic, and transport properties. It may also provide a mean to maximize the coupling with the functional groups of DNA and thus the role of $\gamma_{1/2}$ in Eq. (4), i.e., to increase the DNA reading capability. In Fig. 3e, we show the DOS and transmission of ketone-terminated zGNR, which demonstrates that the edge oxidation has increased the work function and accordingly enhanced metallicity at $E_F$, which makes it a better electrode material than hydrogenated zGNR (Jeong et al. 2013).

In addition to the modification of bulk properties, functionalization of carbon nanoelectrodes at their edges can induce even more drastic effects. For example, whereas two pristine capped CNT electrodes (Mode I-B) generate negligible tunneling currents in spite of the metallic nature of bulk (5,5) CNT (Fig. 3f), the capped CNTs doped by one nitrogen atom at their cap ends can show much increased tunneling currents (Fig. 3g). We have also shown that the charged defect states can provide a much enhanced chemical sensitivity (Kim et al. 2014).

Finally, it should be noted that the unique and excellent coherent charge transport properties of carbon nanomaterials (Figs. 2a—2e) (Charlier et al. 2007; Cresti et al. 2008; Das Sarma et al. 2011; Kang et al. 2009) allow novel device configurations such as Mode II-A and Mode II-B. A significant advantage of Mode II over Mode I is that the magnitude of current signals in Mode II is much higher than that of tunneling currents in Mode I. We will discuss these aspects in Secs. 4 and 5.

## 3. DNA sequencing based on metal electrodes

The idea to read different nucleobases in a ssDNA using tunneling currents can be traced back to the invention of STM, as the DNA molecules were one of the first target of STM applications (Binnig and Rohrer 1984) (Mode I-A). However, due to the significant technical difficulties in sample preparation and reproducibility (Clemmer and Beebe 1991), it took over two decades to make true progress (He et al. 2007; He et al. 2008b; Jin et al. 2009; Ohshiro and Umezawa 2006; Shapir et al. 2008; Tanaka and Kawai 2009). In this case, as in typical STM experiments a ssDNA is placed on an appropriate substrate in ultrahigh vacuum and the STM tip will be moving over the DNA molecule to complete the reading. In principle, the high resolution achievable in STM can allow single-base reading of ssDNA. In practice, however, there exist many technical challenges (Clemmer and Beebe 1991; Tanaka and Kawai 2009). As a way to enhance the STM signal, it was suggested that an STM tip functionalized with one of the four DNA bases could enhance hydrogen bond-mediated interactions between DNA bases or the chemical contrast in STM images (He et al. 2007; Ohshiro and Umezawa 2006). In a slightly different approach yet in the same spirit, it was suggested to employ guanidinium ions that form specific hydrogen bonds to phosphate groups in ssDNA (He et al. 2008b). However, still, the Mode I-A configuration has several disadvantages in view of practical applications: Because the usage of STM requires ultra-vacuum environment, the portability of the DNA sequencer will be quite limited. Furthermore, the interpretation of STM contrast is apparently too challenging to be routinely used for clinical practice.

The transverse tunneling current DNA sequencing approach (Mode I-B) is different from STM-based methods (Mode I-A) in that two nano-scale metal electrodes are now fixed in an electrochemical environment and an ssDNA translocates through the gap between these electrodes. Since the DNA molecules cannot be anchored or the molecular configurations are not well-defined in Mode I-B, it was early on pointed out that tunneling currents can be determined by various factors such as the chemical connectivity between DNA and electrodes, the size and orientation of nucleobases, and the energetic location of their HOMO and lowest-unoccupied molecular orbital (LUMO) (See discussions in Sec. 2.2). In this context, we note that there was an interesting controversy even on which molecule gives the highest conductance (Lagerqvist et al. 2007a; Zikic et al. 2006, 2007; Zwolak and Di Ventra 2005). One of the first proposals of the transverse tunneling current DNA sequencing (Zwolak and Di Ventra 2005) designated dAMP as the highest-signal molecule (energy level location as the major factor) and claimed that the conductance ordering is robust even after thermal fluctuation and environmental effects are taken into account (Krems et al. 2009; Lagerqvist et al. 2006, 2007b). The semi-empirical TB computation level and utilization of a very large bias voltage in this work were criticized in Ref. (Zikic et al. 2006), which identified dGMP as the originator of the largest signal (molecular size as the key element that reduces the vacuum gap and enhances the non-resonant tunneling). Note that the gap size between two tunneling electrodes is in fact an adjustable parameter, which could modify the factor that determines the conductance ordering.

In this context, there was a proposal to use two sets of Au electrodes with different gap sizes and their cross arrangement for simultaneous reading (Bagci and Kaun 2011).

With the Au electrodes, the first two proof-of-principles experiments successfully demonstrated the possibility of transverse-tunneling-current DNA sequencing in Mode I-B. While there still remain tremendous experimental challenges such as the incorporation of a nanopore, etc., these reports represent important milestones in the next-generation DNA sequencing and we here briefly summarize them. The first group employed the Au mechanically controllable break junction electrodes in the Mode I-B configuration and successfully distinguished all four types of DNA bases with the conductance signal order of dGMP > dAMP > dCMP > dTMP (Fig. 4a) (Ohshiro et al. 2012; Tsutsui et al. 2010). As emphasized earlier, the large variations in molecular orientation and resulting conductance in the course of DNA translocation substantially reduce the selectivity, which raise the necessity to control the configuration of DNA nucleobases. Following the STM tip functionalization idea (Mode I-A) (Ohshiro and Umezawa 2006), functionalizing the electrodes in Mode I-B would be also an appealing approach. Note that, however, a functionalization that induces too strong bonds with nucleobases is apparently not an option for DNA sequencing, and instead one needs appropriately weak bonds. Namely, the appropriate choice of functionalization group is crucial for the success of such a scheme. Overcoming this challenge, the second group functionalized the substrate and STM Au electrodes (Mode I-A configured in the fashion of Mode I-B) using mercaptobenzamide and reported the conductance ordering of dAMP > dGMP > dCMP (Fig. 4b) (Huang et al. 2010). The mercaptobenzamide functional groups with two hydrogen-bond donor sites (nitrogen atoms) and one hydrogen-bond acceptor site (carbonyl oxygen) were suggested to establish the desired sliding contacts for the translocating target DNA molecule.

## 4. DNA sequencing based on carbon nanotubes

### 4.1 Mode I-B based on carbon nanotube electrodes

By mid-2000s, CNTs have been extensively studied as the material for nanoelectronics, particularly molecular sensors. Experimentally, with the progress made in the synthesis of CNTs, CNT electrodes with the diameter that provide the single-nucleobase resolution, i.e. < ~ 7.4 Å, could be straightforwardly prepared. In this context, CNTs were theoretically considered as a nanoelectrode material for DNA sequencing. Ref. (Meunier and Krstić 2008), e.g. considered several CNT end atomic structures (H-termination, cap, N-termination) and particularly suggested that nitrogen terminated armchair CNT ends whose lone pairs are exposed toward the target base can provide drastically enhanced tunneling currents and at the same time an enhanced sensitivity in distinguishing a purine from a pyrimidine (Fig. 5a). Even though the capped CNT ends were predicted to be insensitive at the large gap distance of 15 Å (Meunier and Krstić 2008), it was pointed out that they can dramatically enhances the connectivity between DNA molecule and CNT electrodes at short gap distances of 6.4 Å by maximizing π-π interactions (Fig. 5b) (Chen et al. 2012). This device configuration was also suggested to slow down the ssDNA translocation speed. Finally, we have demonstrated that doping capped CNT ends with nitrogen atoms can maximize the role of $E_0$ [denominator in Eq. (4)] at short gap distances or $\gamma_{1/2}$ [numerator in Eq. (4)] at long gap distances, and leads to two completely different sequencing protocols (Fig. 5c) (Kim et al. 2014). Particularly, we have shown that the novel sequencing mode of the latter large-gap distance case results from the different number and arrangement of chemical functional groups in the four nucleobases (C=O, $-NH_3$, and $-CH_3$) and the ability of the substitutionally-doped nitrogen in $sp^2$ carbon network to act both as an electron acceptor and a donor. We finally point out that the strong charge transfer between nucleobase functional groups and N dopant atoms within the CNT caps might also result in the slowdown of translocation speed of ssDNA and better-defined nucleobase configurations within the electrode-electrode gap.

## 4.2 Other modes based on carbon nanotube electrodes

Other than the advantage of its dimension, as a quasi-one-dimensional π-conjugate system, CNT also allows the strong adsorption of ssDNA on its cylindrical wall. Considering CNT as an underlying substrate, it was theoretically proposed that nucleotides adsorbed on a CNT could be identified by measuring their LDOS with an STM probe (Mode I-A) (Meng et al. 2006). Another distinctive and advantageous feature of CNT is its excellent transport properties including its electronic sensitivity to the level of point defects or point functionalization (Goldsmith et al. 2007). Utilizing this property, the possibility of electrical sequencing by introducing single point defect and taking the benefit of defect-dominated conductance in CNT via mode II-A has been proposed (Chen et al. 2013). Going further in terms of geometrical manipulation of CNT, a hollow torus made of CNT that is connected to two CNT electrodes (Mode II-B) has been proposed as a potential sequencing platform (Sadeghi et al. 2014a).

# 5. DNA sequencing based on graphene and graphene nanoribbons

## 5.1 Mode I based on graphene and graphene nanoribbon electrodes

### 5.1.1 Graphene and graphene nanoribbon electrodes

In the Mode I-A configuration, DNA nucleobases adsorbed on graphene was suggested to exhibit distinct STM signals originating from their frontier orbitals (Ahmed et al. 2012). However, as discussed earlier, the STM approach has many practical drawbacks such as the requirement of ultra-high vacuum condition. Being an one-atom-thick 2D materials, it would be natural to consider graphene as nanoelectrodes for DNA sequencing in Mode I-B (Fig. 6a) (Postma 2010). For the graphene nanogap with armchair edges, it was shown that the conductance ordering is determined by the HOMO levels of the four nucleobases and their physical size and orientation with respect to the electrodes govern the signal width (Sec. 2.2) (Prasongkit et al. 2011). As typical in theoretical studies of graphene, graphene edges along the nanogap were terminated by hydrogen atoms. This simplest hydrogenated graphene was proposed to establish hydrogen bonding with DNA and enhance the sensitivity of the graphene DNA sequencer (He et al. 2011). Enhanced graphene-DNA coupling could restrict DNA conformation and accordingly reduce translocation speed. Meanwhile, zGNRs with non-hydrogenated armchair edges were also proposed as DNA sequencing electrodes in Mode I-B (Zhang et al. 2014).

### 5.1.2 Functionalized graphene and graphene nanoribbon electrodes

Rather than the pristine hydrogen-terminated graphene, it would be in general more realistic to consider doped and/or defected graphene. Considering the wet device processing and electrochemical operating conditions, it would be particularly appropriate to consider oxygen-functionalized graphene. In Sec. 2.3, we have already pointed out that oxygen edge functionalization can in principle enhance the metallicity of GNR electrodes. In addition to the higher current carrying capacity, another prerequisite of good sensing electrodes would be high molecular selectivity. We have envisioned that exploiting the additional degree of freedom in functionalization along the gap-side edge of graphene could lead to better selectivity for graphene-based DNA sequencer. After confirming that the tunneling currents are dramatically increased for the ketone-terminated zGNR, we next considered alternating OH- and H-functionalization of the gap-side armchair edge of zGNR and showed that it not only enables the extension of $\pi$-conjugation within GNR to DNA nucleobases but also can result in destructive and Fano quantum interference patterns (Fig. 6b) (Jeong et al. 2013).

Given its experimentally proved viability in the Au electrodes (Sec. 3) (He et al. 2007; He et al. 2008b; Huang et al. 2010; Ohshiro and Umezawa 2006), it was also proposed to functionalize the edge of graphene electrodes with a guanidinium ion that grab phosphate groups and a reader nucleobase that induces stronger hydrogen bonds with target nucleobases (Prasongkit et al. 2013).

## 5.2 Mode II based on graphene and graphene nanoribbons

### 5.2.1 Graphene nanopore (Mode II-B)

Due to its two-dimensional (2D) membrane nature, drilling a hole within graphene can in principle provide us with a single-material nanopore and electrical current measuring platform (Mode II-B). Indeed, DNA translocation through graphene nanopore was experimentally demonstrated by the longitudinal ionic current modulation (Garaj et al. 2010; Merchant et al. 2010; Schneider et al. 2010), and more recently the simultaneous measurement of longitudinal ionic and transverse electrical current modulations has been reported (Traversi et al. 2013). Although the actual sequencing of ssDNA still remains to be demonstrated, these experimental advances definitely represent the significant advances made toward the graphene-based DNA sequencing.

In terms of the reading mechanism, we remind the readers that the major signals in Mode II originate from the current flow through the inorganic channel material, i.e. graphene itself and the presence of DNA nucleobases appear as perturbing signals. For the GNRs, as discussed in Sec. 2.3, we have an added complexity of edge effects, which particularly in the zGNRs ferromagnetically spin-polarized and become the major electron transport channels. Ref. (Nelson et al. 2010) considered a semiconducting aGNR channel with a nanopore and claimed that the larger purine G and A will produce higher conductance than smaller pyrimidine C and T. It was also suggested that such a device configuration gives the possibility of orientation-independent sequencing. While the presence of a nanopore reduces the conductance of the GNR electrode, it was also suggested that shaping the nanoribbon geometry with the lateral constriction integrated with the motion of charged group and screening depending on electrical gate or external charge could enhance the sensitivity of the DNA sequencer (Girdhar et al. 2013, 2014).

Remind that the important difference between aGNRs and zGNRs is that the central and edge regions become the main transport channels, respectively (Sec. 2.3). Given that the conductance signals can be significantly reduced in aGNRs by the presence of a nanopore, one can instead consider zGNR as a potentially better DNA sequencing platform (Fig. 6c) (Saha et al. 2011). It was also suggested that decorating the nanopore edge with nitrogen atoms can significantly enhance the conductance magnitude. In addition, it was also claimed that the zGNR-nanopore platform could be used to detect DNA methylation (Ahmed et al. 2014b). However, a more recent study has pointed out some critical issues in zGNR nanopore-based sequencing (Avdoshenko et al. 2013). First, although the nucleobase orientation-independent sequencing is often regarded as an important advantage of the nanopore-based approach (Avdoshenko et al. 2013), it was shown that the current signals are in fact highly dependent on DNA orientation. Second, although it enhances the magnitude of electric currents, the edge-dominated current in zGNR could mask the difference in nucleobase signals. Similarly, considerable dependence of conductance on the nucleobase orientation and adjacent nucleobases has been pointed out (Shenglin et al. 2014). Still, it was suggested that such difficulties might be overcome by employing a novel signal analysis approach which statistically utilizes time-dependent current spectrum and the correlation function of conductance profiles obtained by multilayer nanopore device (Ahmed et al. 2014a).

5.2.2 Fano resonance in graphene nanoribbon electrodes (Mode II-A)

Another idea involving GNR is to utilize the Fano-resonance signals appearing when the nucleobases are adsorbed on the aGNR electrode (Fig. 6d) (Min et al. 2011; Rajan et al. 2014). It was claimed that the single nucleobase interacting with GNR via π-π interactions will result in the characteristic conductance dips originating from the nucleobase HOMOs at nucleobase-specific gate bias voltages. While the idea is novel in both device geometry and sensing mechanism, the requirement of picking up quantum interference patterns at large yet well-defined gate bias voltages will certainly make its experimental realization a very difficult task. It was still suggested that using the data-mining technique and two-dimensional transient autocorrelation functions could enable the detection of such Fano-resonance signals.

## 6. DNA sequencing based on other 2D materials and their van der Waals heterostructures

Following the great success of research centered on graphene, novel monolayer 2D crystals beyond graphene such as hBN, silicene, molybdenum disulphide ($MoS_2$), and other transition metal dichacogenides, etc., and their multilayer structures have been attracting great attention. As the natural extension of the DNA sequencing based on monolayer graphene, there appeared a number of reports that considered the potential of such graphene-like materials as the DNA sequencer (Fig. 7).

In two independent studies (He ; Sadeghi et al. 2014a), it was proposed to utilize a nanopore formed within the graphene bilayer. In these device configurations, the top and bottom graphene layers will function as separate electrodes. However, once more, different opinions were derived: While Ref. (He) suggested that inter-graphene-layer tunneling currents are negligible and the bridging nucleobases provide the major tunneling conductance of the G > A > T > C ordering (Mode I-B), Ref. (Sadeghi et al. 2014a) claimed that inter-layer tunneling currents will be relatively large in the beginning and nucleobases that pass through the pore will provide perturbing signals (Mode II-B). In another study, graphene/hBN/graphene vertical heterojunctions were considered in light of the inter-graphene-layer tunneling currents (He et al. 2014). It was interestingly found that, unlike the ABA stacking case, the ABC counterpart will result in the suppression of tunneling currents or the background noise and thus much amplified vertical transmissions mediated by the molecule inserted in a nanopore (from Mode II-B to Mode I-B).

A single-layer $MoS_2$ were shown to be a potentially versatile DNA sequencer that can operate in both longitudinal ionic and transverse electric current modes (Fig. 7a) (Farimani et al. 2014). It was suggested that a nanopore within a $MoS_2$ monolayer will provide the ionic current ordering of dAMP-dTMP > dGMP > dTMP > dCMP-dGMP > dCMP for the translocating double-stranded DNA and the Mode II-B transverse current ordering of G > A > C > T according to the induced DOS due to the DNA bases placed within the nanopore. The same conductance ordering was suggested for the armchair $MoS_2$ nanoribbon according to the adsorption strength and bandgap modulation upon adsorption of nucleobases (Mode II-A).

A silicene monolayer was also considered in several studies. In Ref. (Amorim and Scheicher 2014), the authors claimed that Si atoms will interact more strongly with the ketone groups, so the single-ketone-group-containing C and G (see Fig. 2) will adsorb on the buckled silicene atoms more strongly than A and T. It was suggested that these behavior will result in detectable contrasts in both STM images (Mode I-A) and gate-modulated conductance (Mode II-A). An independent study has shown that silicene nanopore can also act as the Mode II-B sequencer (Fig. 7b) (Sadeghi et al. 2014b). Also, in Ref. (Thomas et al. 2014), hBN, silicene, and $MoS_2$ nanoribbons were considered for the Mode II-A sequencer, and hBN was identified as a promising material that can produce large Fano resonance signals (Fig. 7c).

## 7. Issues in computational studies

Because of the complexity of experimental conditions in the targeted transverse electrical current DNA sequencing, most theoretical and computational reports had to resort to simplification in certain aspects of modeling and simulation. Some of the adopted assumptions in those studies might be too drastic, which requires one to judge their reliability with special care. In this section, we discuss these potential missing elements and point out desirable directions in the future theoretical study.

*Calculation levels*. It should be reminded that from the beginning of theoretical investigations there were controversies on the reliability of computational predictions in terms of computational levels (Sec. 2.1). The first report that utilized a TB scheme for the Au electrodes in Mode I-B predicted that dAMP will produce the largest signal (Lagerqvist et al. 2007a; Zwolak and Di Ventra 2005). On the other hand, a similar study that employed the first-principles DFT computational level identified dGMP as the source of the largest signal (Zikic et al. 2006, 2007), which was later found to be in agreement with experimental results (Ohshiro et al. 2012; Tsutsui et al. 2010). We mentioned that similar controversies exist on the reliability of computational levels in the studies carried out for the GNR (Sec. 5.2.1) (Avdoshenko et al. 2013; Chang et al. 2014; Nelson et al. 2010; Saha et al. 2011) and bilayer graphene (Sec. 6) (He et al. 2012 ; Sadeghi et al. 2014a) sequencers.

*Modeling*. Given that the electrode (probe) and nucleobases (target) should be the key elements that determine the current flow through an electrical current DNA sequencer, most computational studies have so far omitted in the quantum transport calculations neighboring nucleobases and/or the backbone as well as complex environments such as water and counterions. First, on the omission of neighboring nucleobases, a recent TB study claimed that random displacement of neighboring nucleotides might cause large error rates Ref. (Alvarez et al. 2014). Another study showed that the effect of neighboring nucleobase can be even more significant for the zGNR nanopore device using first-principles scheme (Shenglin et al. 2014).

Next, complex environments can in principle modify electrical currents calculated in vacuum by providing dielectric and Debye screening (by water and mobile counterions, respectively) and compensating negative phosphate backbone charges (by $Na^+$ or $K^+$ that condensate directly onto them). While a series of early TB studies concluded that such environmental effects should be negligible (Krems et al. 2009; Lagerqvist et al. 2006, 2007b), a more recent first-principles study on aGNR nanopore claimed that explicitly negatively charged $PO_4$ models will provide conclusions on selectivity that are different from those obtained with neutral models (Avdoshenko et al. 2013).

More importantly, liquid environments can affect the thermal and structural fluctuations of DNAs. These will not only significantly affect the DNA translocation dynamics (Fyta et al. 2011; Luan et al. 2012; Zwolak and Di Ventra 2008), but should also result in the inherently statistical spread of the measured electrical currents. To describe such molecular fluctuation effects on electrical currents, one often resort to the multiscale computational approach that is standard in molecular electronics research (Kim and Kim 2010; Kim et al. 2005): One first carry out molecular dynamics (MD) simulations based on classical force fields (FF), and next compute transmissions for the relevant MD snapshots (He et al. 2011; He et al. 2012; Krems et al. 2009; Lagerqvist et al. 2006, 2007b; Min et al. 2011; Sadeghi et al. 2014a; Zikic et al. 2006). Reminding the above-discussed controversies on the computational levels, then one might suspect the validity of MD snapshots derived from these non-*ab initio* FFMD runs.

Finally, in Mode I-B, one can also question whether the coherent NEGF formalism, i.e. Eq. (3), is sufficient to describe the true nature of charge transport. Given that in the DNA sequencer nucleobases are not directly linked with the electrodes through covalent bonds (unlike in the conventional molecular electronic junctions), there can in principle exist strong contributions to electrical currents that originate from incoherent electron hopping or sequential single electron tunneling. Taking this weak-coupling Coulomb blockade regime, or considering DNA molecule as quantum dots, Ref. (Guo et al. 2012) concluded that the DNA sequencing should be still possible independent of molecular orientation and positions. It would be still desirable to treat the co-tunneling and sequential electron tunneling contributions on an equal footing and check the validity of such conclusions.

## 8. Outlook

The measurement of transverse electrical currents represents a promising DNA sequencing approach that can achieve single-molecule resolution without labeling and amplification. In its development, theory and computation have been playing an important role. Not to mention the original idea, most proposals that involve carbon and other low-dimensional nanomaterials have been first put forward in theoretical studies. Considering the complexity of experiments, this trend can be expected to continue in this field.

However, it is also clear much more needs to be done to realize the complete and reliable simulation of transverse electrical current DNA sequencing. We pointed out that there exist potentially critical issues in previous computational studies, e.g. missing elements in modeling and/or insufficient simulation levels (omission of backbone, counterions, solvents, realistic electrode *and* molecular geometries, thermal fluctuations, incoherent hopping process, etc.), which resulted in differing conclusions on basically identical device systems. This naturally points us toward desirable future research directions. One example would be incorporating accurate descriptions of complex environmental effects into a first-principles charge transport calculation method that also includes incoherent charge hopping processes.

**Acknowledgements**

This research was supported by Global Frontier Program (No. 2013M3A6B1078881) of the National Research Foundation (NRF) funded by the Ministry of Science, ICT and Future Planning of Korea. HSK was additionally supported by the NRF 2013-Global Ph.D. Fellowship Program.

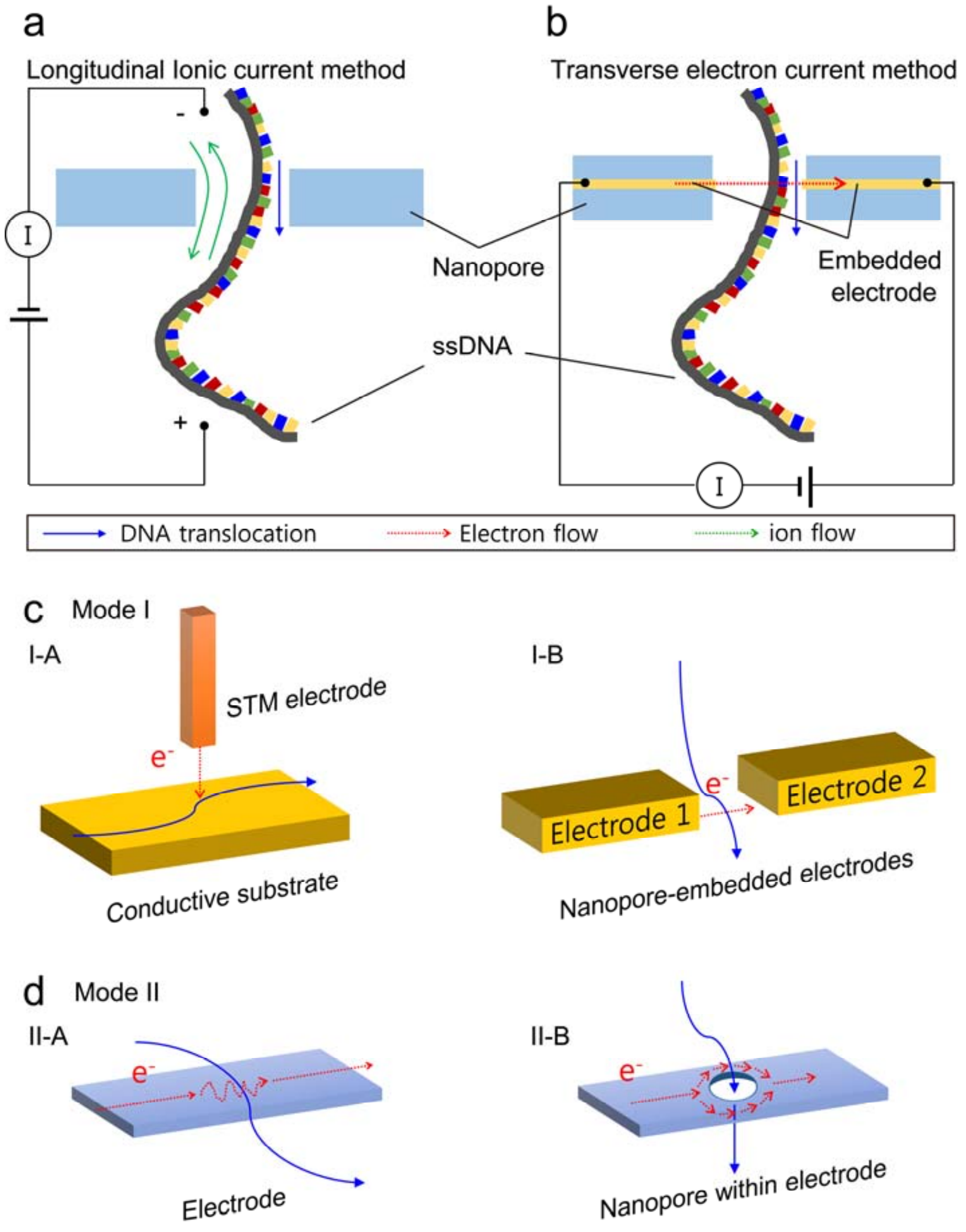

**Fig. 1.** Schematics of device configurations for the third-generation DNA sequencing based on (a) longitudinal-direction ionic current and (b) transverse-direction electron current. Categorization of the transverse current DNA sequencers proposed in the literature based on the device configurations and sensing mechanisms: (c) Mode I includes the STM approach (I-A) and the tunneling nanogap (I-B), (d) Mode II consists of a nanochannel where ssDNA temporarily adsorbed during the translocation process (II-A) and a nanochannel in which a nanopore is embedded (II-B).

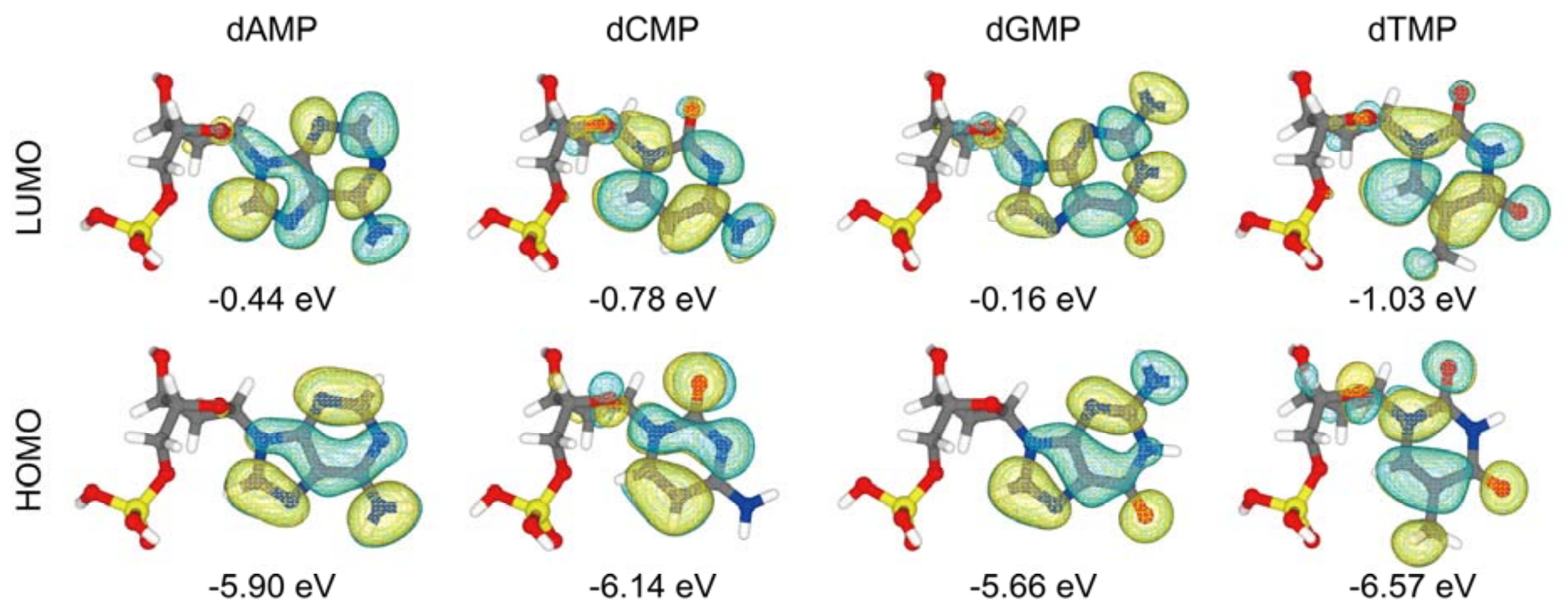

**Fig. 2.** Atomic structures and frontier orbitals (HOMO and LUMO) of dAMP, dCMP, dGMP, and dTMP. Corresponding energy eigenvalues taken from (Ohshiro et al. 2012) are shown together.

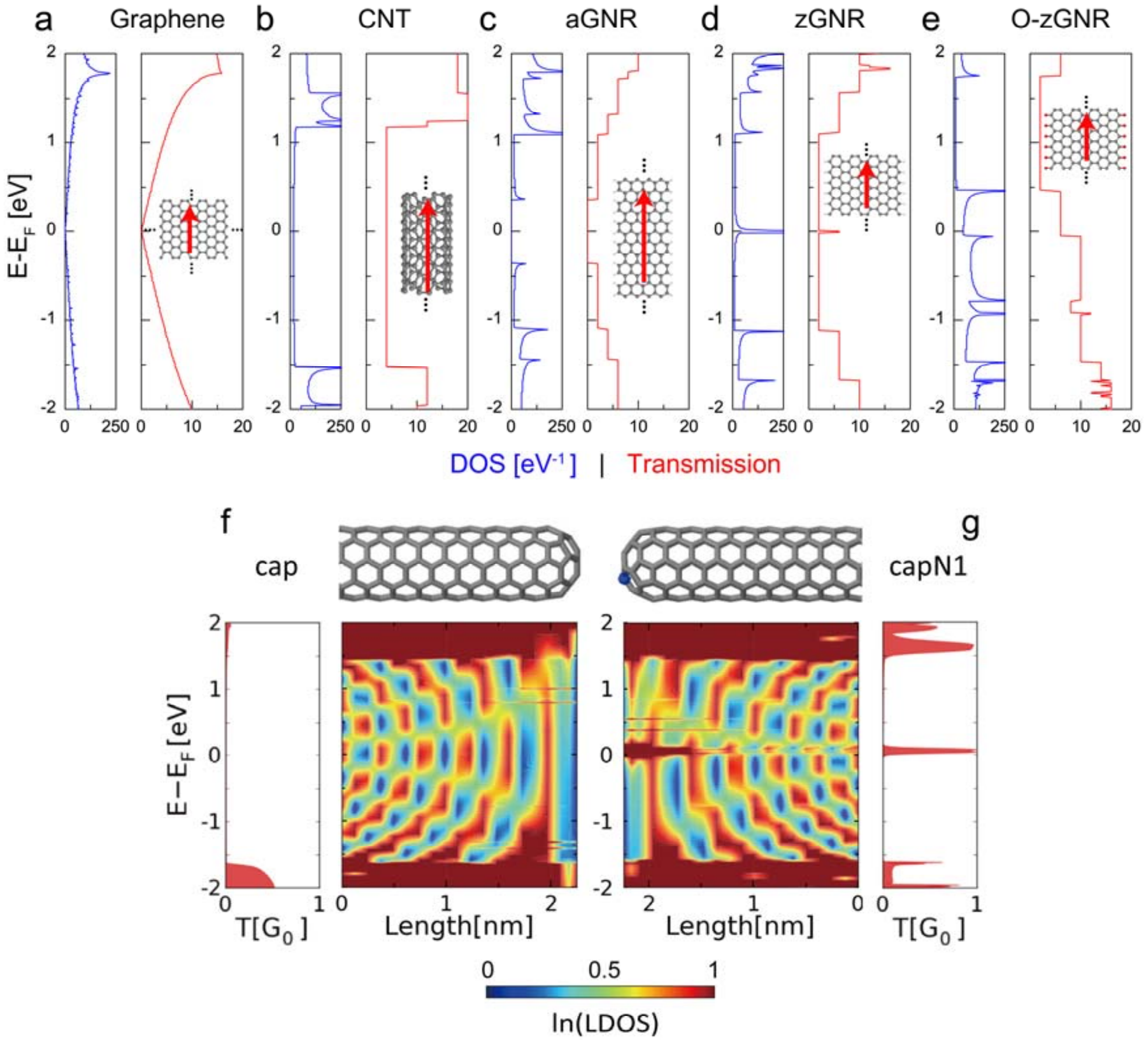

**Fig. 3.** DOS and transmission functions (calculated within GGA at the DZP basis set level) of low-dimensional carbon nanoelectrodes utilized as DNA sequencers: (a) graphene, (b) CNT, (c) aGNR, (d) hydrogenated zGNR, and (e) ketone-terminated zGNR. Corresponding atomistic models are shown in the insets of transmission panels, and red arrows indicate the electron transport directions in each electrode. LDOS, transmission functions, and atomistic models of (f) capped CNT and (g) N-doped capped CNT. Figures reproduced with permissions from: (f-g) Ref. (Kim et al. 2014),© Wiley-VCH Verlag GmbH & Co. KGaA, Weinheim.

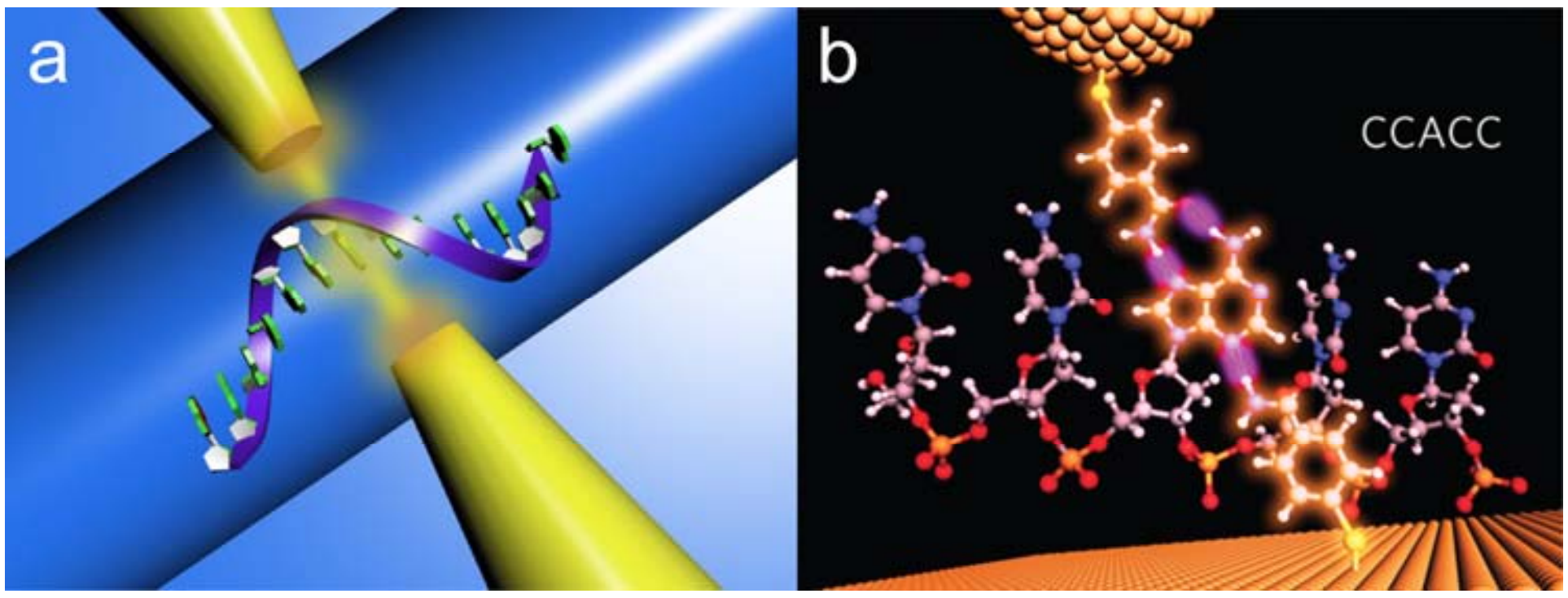

**Fig. 4.** Device geometries based on (a) bare and (b) functionalized Au nanogap electrodes. Figures reproduced with permissions from: (a) Ref. (Ohshiro et al. 2012; Tsutsui et al. 2010), © Nature Publishing Group; (b) Ref. (Huang et al. 2010), © Nature Publishing Group.

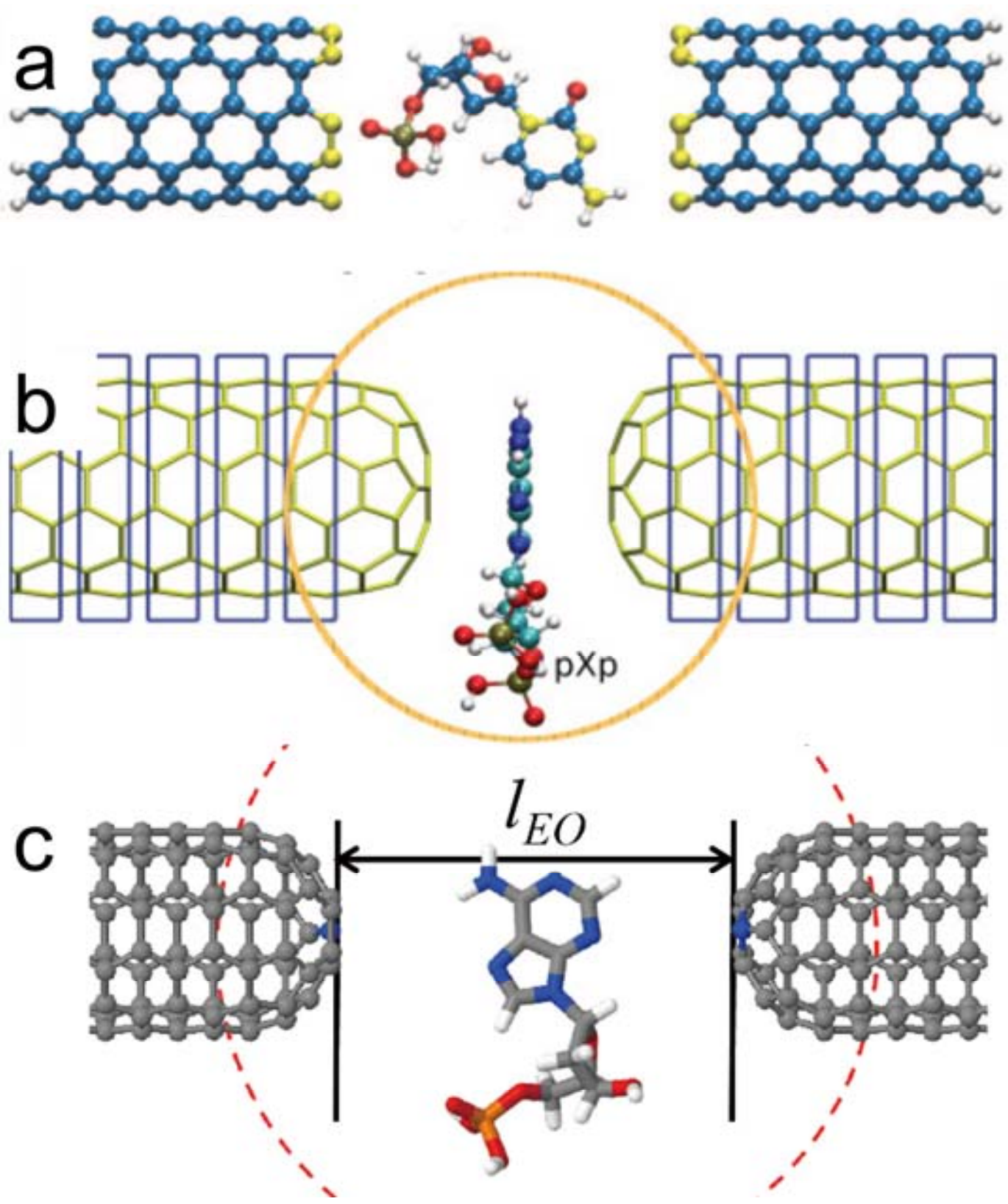

**Fig. 5.** Device geometries based on (a) nitrogen-decorated open-ended, (b) capped, and (c) nitrogen-doped capped CNT electrodes. Figures reproduced with permissions from: (a) Ref. (Meunier and Krstić 2008), © American Institute of Physics; (b) Ref. (Chen et al. 2012),© American Physical Society; (c) Ref. (Kim et al. 2014),© Wiley-VCH Verlag GmbH & Co. KGaA, Weinheim.

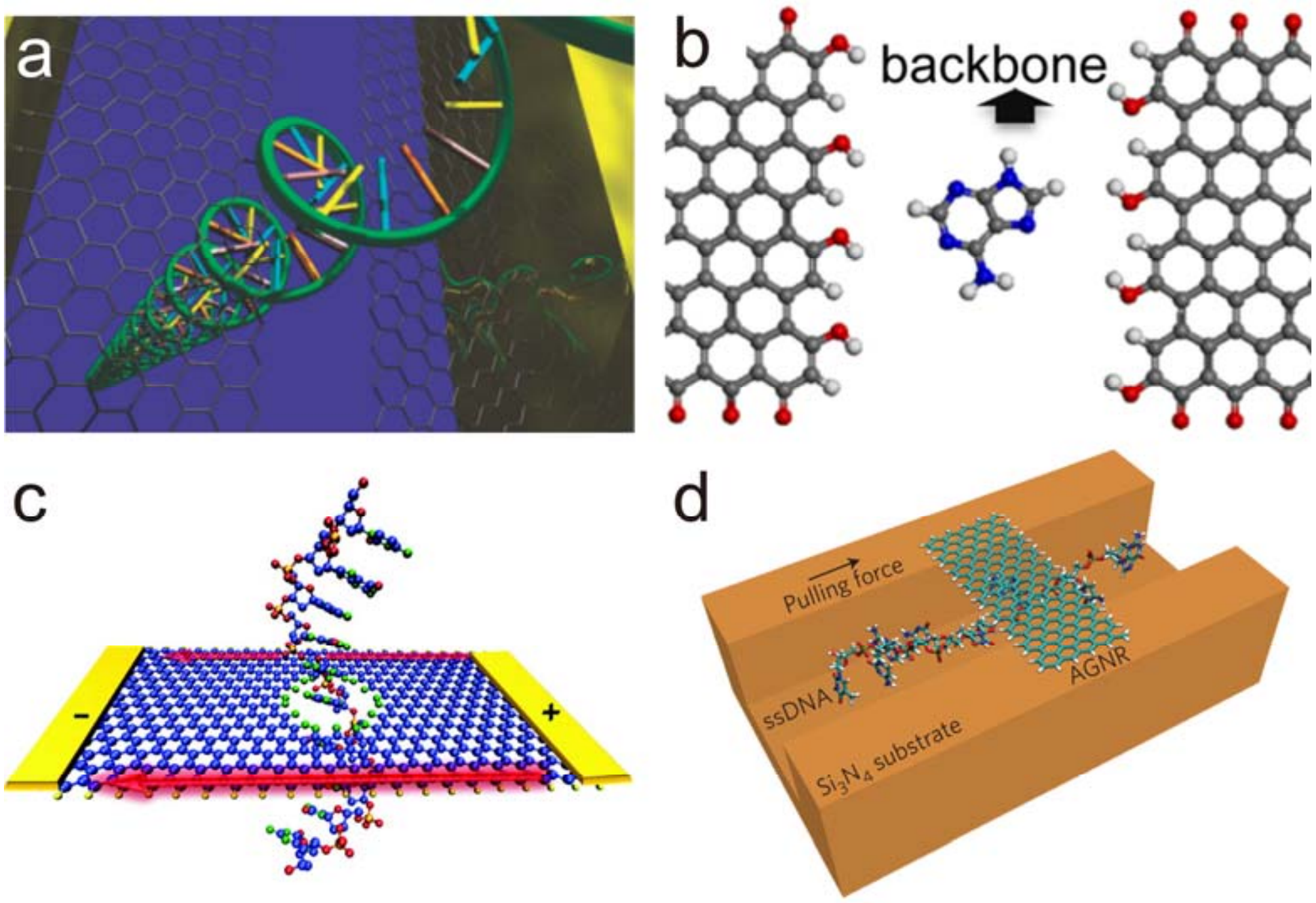

**Fig. 6.** Device geometries using graphene and GNR-based electrodes. (a) Illustration for the graphene nanogap device. (b) Oxygen-functionalized GNR nanogap, (c) nitrogen-dope GNR nanopore, and (d) GNR nanochannel devices. Figures reproduced with permissions from: (a) Ref. (Postma 2010), © American Chemical Society; (b) Ref. (Jeong et al. 2013),© AIP Publishing LLC; (c) Ref. (Saha et al. 2011), © American Chemical Society; (d) Ref. (Min et al. 2011),© Nature Publishing Group.

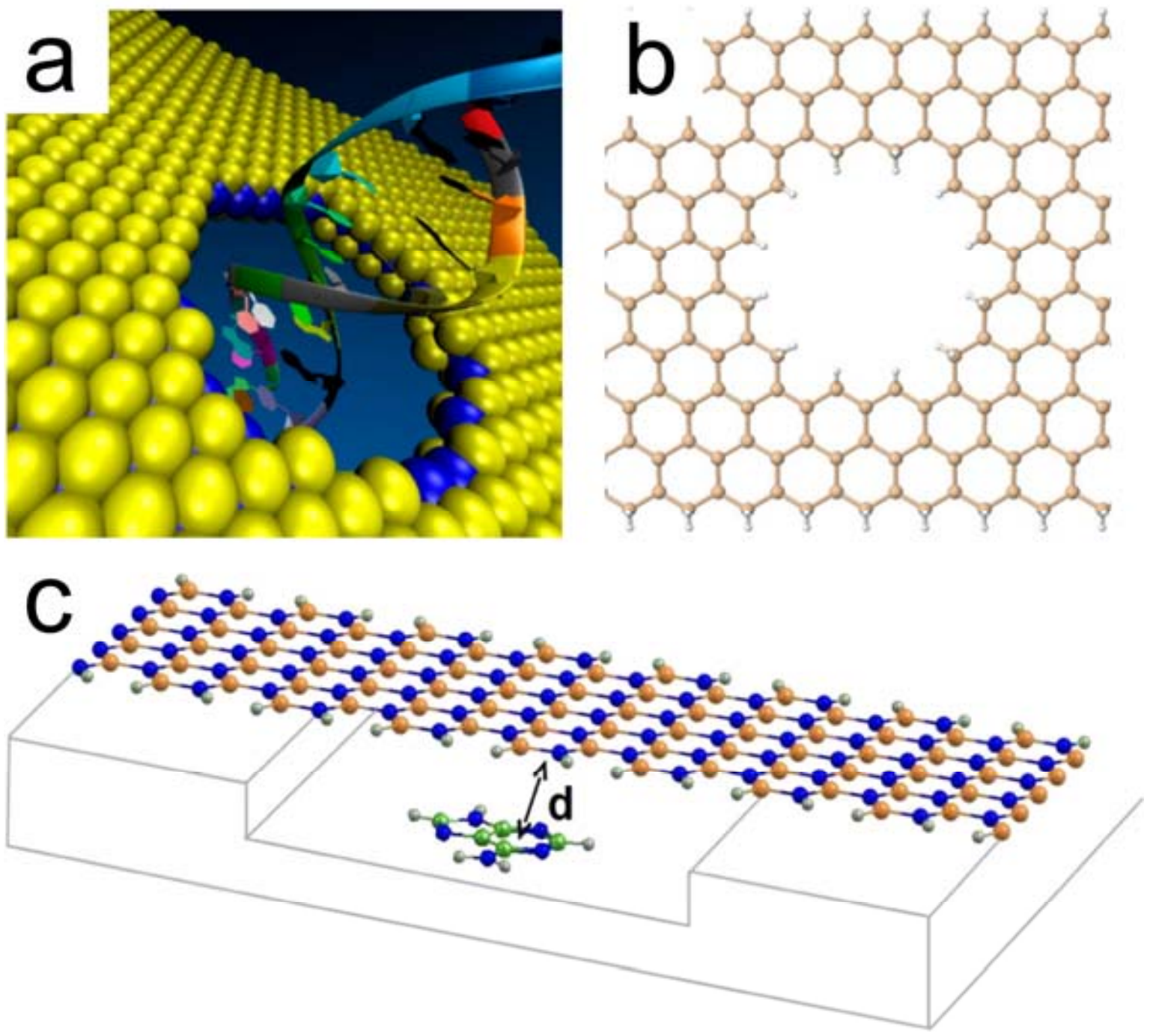

**Fig. 7.** Device geometries using two-dimensional materials beyond graphene. (a) $MoS_2$ nanopore, (b) silicene nanopore, and (c) hBN nanochannel. Figures reproduced with permissions from: (a) Ref. (Farimani et al. 2014),© American Chemical Society; (b) Ref. (Sadeghi et al. 2014b), © AIP Publishing LLC; (c) Ref. (Thomas et al. 2014),© American Chemical Society.

**Table 1.** Transverse electrical current DNA sequencing devices proposed in the literature classified in terms of electrode materials, device configurations, modeling features, simulation levels, signal type and ranking, and reading mechanism. Device configurations correspond to those of Figures 1C–1D, and *d* represents nanogap distance or pore diameter. Source-drain bias voltage ($V_{SD}$) is specified only when the signal was explicitly obtained at a finite $V_{SD}$. Otherwise, signal rankings are based on the zero-bias conductance values. Abbreviations used in this table: SEL, single energy level model to DFT; GGA, generalized gradient approximation; LDA, local density approximation; DZP, double-ζ-polarized atomic-orbital basis sets; SZP, single-ζ-polarized atomic-orbital basis sets; SZ, single-ζ atomic-orbital basis sets; PW, plane-wave basis sets.

| Material | Device configuration | Modeling | Simulation | Signal type & ranking | Reading mechanism | Ref. |
|---|---|---|---|---|---|---|
| **Au** | I-B<br>*d* = 15 Å | Base + backbone (passivated)<br>Edge-on (backbone-mediated) | TB + NEGF | dAMP > dGMP > dCMP > dTMP<br>($V_{SD}$ = 0 – 0.1 V) | HOMO, LUMO position<br>+ Coupling strength | Zwolak and Di Ventra) (2005) |
| **Au** | I-B<br>*d* = 15 Å | Base + backbone (passivated)<br>Edge-on (backbone-mediated) | DFT + NEGF | dGMP > dAMP > dCMP > dTMP<br>($V_{SD}$ < 0.1 V) | Configuration-dependent reduce in tunneling gap | Zikic et al. (2006) |
| **Au** | I-B<br>(a) *d* = 7 Å<br>(b) *d* = 14 Å | Base + backbone (passivated)<br>(a) Face-on<br>(b) Edge-on (backbone-mediated) | DFT (GGA, DZP)<br>+ NEGF | cross-contact<br>(a) dGMP >> others<br>($V_{SD}$ = 1.2 V)<br>(b) dCMP >> others | Number of peaks in real-time conductance profile, differentiated by number of rings within each base | Bagci and Kaun (2011) |
| **Au (cytosine probe)** | I-B<br>*d* = 21.8 Å | Base + backbone (passivated)<br>Edge-on (backbone-mediated) | DFT (GGA, DZP)<br>+ NEGF | ● dAMP, dGMP > dCMP, dTMP<br>($V_{SD}$ = 0.1 V)<br>● dAMP > dGMP<br>($V_{SD}$ = 0.75 V)<br>● dTMP > dCMP<br>($V_{SD}$ = 0.25 V) | Size of target molecule<br>+ Characteristic Bias-dependent shift of molecular peaks | He et al. (2008a) |
| **Au (cytosine and thiocarbamide)** | I-B | Base + backbone (charged)<br>Edge-on (backbone-mediated) | DFT (LDA, DZP) + NEGF | dAMP > dGMP > dTMP > dCMP | Size of target molecule (coupling) | Pathak et al. (2012) |
| **Au** | I-B<br>*d* = 12 Å | Base<br>Face-on & Edge-on | DFT (LDA, DZP) | G > A ~ C > T<br>(charge state junction point) | Incoherent transport at weak coupling limit.<br>Energy eigenvalue-dependent molecular fingerprint | Guo et al. (2012) |
| **CNT (substrate)<br>+ Metal (tip)** | I-A | Base + sugar | DFT (LDA, PW) | - | Molecular DOS | Meng et al. (2006) |
| **CNT (end N doping)** | I-B<br>*d* = 15 Å | Base + backbone (passivated)<br>Edge-on (backbone-mediated) | DFT (GGA, DZ) + NEGF | dAMP, dGMP > dCMP, dTMP | Coupling by the number of rings within the molecule | Meunier and Krstić (2008) |
| **CNT (cap)** | I-B<br>$d$ = 6.4 Å | Base + backbone (passivated)<br>Face-on | DFT (LDA , DZP)<br>+ NEGF | dAMP > dGMP > dCMP > dTMP | Functional group-dependent coupling | Chen et al. (2012) |
| **CNT (cap N doping)** | I-B<br>(a) $d$ = 6.5 Å<br>(b) $d$ = 12 Å, 14 Å | Base + backbone (passivated)<br>(a) Face-on<br>(b) Edge-on | DFT (GGA, DZP)<br>+ NEGF | Dual mode<br>(a) dGMP > dAMP > dTMP > dCMP<br>(b) dTMP > dGMP > dCMP > dAMP | (a) HOMO-level location (Face-on)<br>(b) Functional group-dependent characteristic charge transfer (Edge-on) | Kim et al. (2014) |
| **Graphene (substrate) + Metal (tip)** | I-A | Base | DFT (LDA, PW) | - | Molecular LDOS | Ahmed et al. (2012) |
| **Graphene (H-gap edge)** | I-B<br>$d$ = 14.7 Å | Base + backbone (passivated)<br>Edge-on, 180° rotation | DFT (GGA, SZP) + NEGF | dGMP > dAMP > dCMP > dTMP<br>($V_{SD}$ = 1 V) | HOMO level location<br>+ Size (Purine > Pyrimidine) | Prasongkit et al. (2011) |
| **zGNR (no edge passivation)** | I-B<br>$d$ = 11 Å | Base<br>Edge-on | DFT (GGA) + NEGF | G > A > T > C<br>($V_{SD}$ = 0 ~ 1 V) | Difference in chemical composition and structures | Zhang et al. (2014) |
| **Graphene (guanidinium ion & cytosine reader)** | I-B<br>$d$ = 23.82 Å | Base + backbone (charged)<br>Edge-on (backbone-mediated) | DFT (GGA, SZP) + NEGF | dGMP > dAMP > dTMP > dCMP<br>($V_{SD}$ < 0.4 V) | HOMO level<br>Bias dependent shift of transmission resonance peaks based on molecular species and coupling strength | Prasongkit et al. (2013) |
| **zGNR (O for zigzag edge, H/OH for armchair gap edge)** | I-B<br>$d$ = 12 Å | Base<br>Edge-on, 60° rotation | DFT (GGA, DZP)<br>+ NEGF | Constraint-angle<br>G >T ≳ C ≳ A | Geometry (Size)<br>+ Destructive quantum interference | Jeong et al. (2013) |
| **Graphene bilayer lateral contact** | I-B<br>$d$ ~ 3.5 Å | Base<br>Edge-on | FFMD<br>SEL | G > A > T > C | HOMO level ordering | He et al. (2012) |
| **aGNR + nanopore (H edge)** | II-B<br>$d$ = 14.5 Å | Base | DFT (GGA, PW) | G, A > C, T | Molecular size-dependent interaction and modulation of conductance through nanoribbon | Nelson et al. (2010) |
| **zGNR (H edge) + nanopore (N edge)** | II-B<br>$d$ = 12 Å | Base | DFT (GGA, DZP)<br>+ NEGF | G > C > A > T | Base-specific modulation of the charge density | Saha et al. (2011) |
| **zGNR + nanopore (H edge)** | II-B<br>$d$ = 16 Å | Base + backbone<br>(a) passivated<br>(b) charged | FFMD<br>TB + NEGF | (a) dTMP > dAMP > dGMP > dCMP<br>(Angle-independent)<br>(b) dGMP > dCMP > dTMP > dAMP<br>(Highly angle-dependent) | Charge rearrangement due to charged $PO_4$ group location | Avdoshenko et al. (2013) |
| **zGNR + nanopore (H edge)** | II-B<br>*d* = 12 Å | Base (+ methyl backbone) | DFT (GGA, SZ) + NEGF | C > A > T > G<br>($V_{SD}$ > 0.4 V)<br>(Highly motion-dependent) | Base-specific modulation of the capacity of electrode's original transport channel | Shenglin et al. (2014) |
| **aGNR** | II-A | Base<br>Adsorbed, from MD trajectory | FFMD<br>DFT (GGA, DZP) + NEGF | - | Characteristic conductance dip originated from HOMO level | Min et al. (2011) |
| **Bilayer graphene nanopores** | II-B<br>*d* = 15 Å | Base | DFT (GGA, DZP) + NEGF | T > A > C > G | Nanopore shape- and phosphate deoxyribose-dependent conductance modulation | Sadeghi et al. (2014a) |
| **Torus-containing nanoribbon** | II-B<br>*d* = 16 Å | Base | DFT (GGA, DZP) + NEGF | G > C > T > A | Nanopore shape- and phosphate deoxyribose-dependent conductance modulation | Sadeghi et al. (2014a) |
| **$MoS_2$** | II-A / II-B | Base | DFT (GGA, DZP) | Binding energy & change in band gap<br>G > A > C > T | Interaction strength purine > pyrimidine | Farimani et al. (2014) |
| **Silicene ZNR (H edge) + nanopore (H edge)** | II-B<br>*d* = 17 Å | Base | DFT (GGA, DZP) + NEGF | T > C > G > A<br>($V_{SD}$ = 0.55 V): | Base-specific change in electrical properties of the ribbon | Sadeghi et al. (2014b) |
| **Silicene** | I-A / II-A | Base | DFT (GGA, DZP) + NEGF | – | Molecule-dependent characteristic adsorption geometry and corresponding conductance dip | Amorim and Scheicher (2014) |
| **hBN, Silicene, & $MoS_2$** | II-A | Base | DFT (GGA, DZP) + NEGF | – | Molecule-dependent characteristic conductance dip | Thomas et al. (2014) |